\documentclass[useAMS,usenatbib]{mn2e}
\usepackage{eso-pic,graphicx}

\usepackage[utf8]{inputenc}
\usepackage[normalem]{ulem}
\usepackage{soulutf8} 

\title[Characterisation of HS 2231+2441]
  {HS\,2231+2441: an HW~Vir system composed by a low-mass white dwarf and a brown dwarf 
   \thanks{Based on observations carried out at the Observat\'orio 
   do Pico dos Dias (OPD/LNA), in Brazil and with the William Herschel Telescope at 
   the Observatorio del Roque de los Muchachos.}}
\author[L.~A.~Almeida et al.]
  {L.~A.~Almeida$^{1}$\thanks{E-mail: leonardodealmeida.andrade@gmail.com},
   A. Damineli$^{1}$,
   C.~V.~Rodrigues$^{2}$,
   M.~G.~Pereira$^{3}$ and 
   F.~Jablonski$^{2}$
  \newauthor 
   \\
  $^{1}$Instituto de Astronomia, Geof\'isica e Ci\^encias Atmosf\'ericas\\
   Rua do Mat\~ao 1226, Cidade Universit\'aria S\~ao Paulo-SP, 05508-090, Brazil \\
  $^{2}$Instituto Nacional de Pesquisas Espaciais/MCTIC \\
   Avenida dos Astronautas 1758, S\~ao Jos\'e dos Campos, SP, 12227-010, Brazil\\
  $^{3}$Departamento de F\'isica, Universidade Estadual de Feira de Santana \\ 
   Av. Transnordestina, S/N, Feira de Santana, BA, 44036-900, Brazil \\
  }
\date{Released 2002 Xxxxx XX}

\pagerange{\pageref{firstpage}--\pageref{lastpage}} \pubyear{2011}

\def\LaTeX{L\kern-.36em\raise.3ex\hbox{a}\kern-.15em
    T\kern-.1667em\lower.7ex\hbox{E}\kern-.125emX}

\begin{document}

\label{firstpage}

\maketitle

\begin{abstract}
HW~Vir systems are rare evolved eclipsing binaries composed by a hot compact star and a low-mass main-sequence star in a close orbit. These systems provide a direct way to measure the fundamental properties, e.g. masses and radii, of their components, hence they are crucial to study the formation of sdB stars and low-mass white dwarfs, the common-envelope phase, and the pre-phase of cataclysmic variables. Here we present a detailed study of HS\,2231+2441, an HW Vir type system, by analysing BVR$_{C}$I$_{C}$ photometry and phase-resolved optical spectroscopy. The spectra of this system, which are dominated by the primary component features, were fitted using NLTE models providing effective temperature T$_{\rm eff} =$ 28500$\pm$500\,K, surface gravity $\log g = 5.40\pm0.05~\rm cm\,s^{-2}$, and helium abundance $\log(n(He)/n(H)) = -2.52\pm0.07$. Geometrical orbit and physical parameters were derived by modelling simultaneously the photometric and spectroscopic data using the Wilson-Devinney code. We derive two possible solutions for HS\,2231+2441 that provide component's masses: M$_1 =0.19\,\rm M_{\odot}$ and M$_2 = 0.036\,\rm M_{\odot}$ or M$_1 = 0.288\,\rm M_{\odot}$ and  M$_2 = 0.046\,\rm M_{\odot}$. Considering the possible evolutionary channels to form a compact hot star, the primary of HS\,2231+2441 probably evolved through the red-giant branch scenario and does not have a helium-burning core, which is consistent with a low-mass white dwarf. Both solutions are consistent with a brown dwarf as the secondary. 
\end{abstract}

\begin{keywords}
 binaries: eclipsing $-$ stars: fundamental parameters $-$ stars: individual: HS 2231+2441 $-$ stars: brown dwarfs $-$ white dwarfs.
\end{keywords}

\section{Introduction}

HW~Vir systems are evolved eclipsing binaries with a short orbital period and composed by a hot compact star and a low-mass main-sequence star. In the most plausible scenario to form these systems, the progenitor of the hot compact star evolves faster than the secondary (a low-mass main-sequence star) starting an unstable mass transfer phase and subsequently a common envelope. During the common-envelope stage, the secondary spirals in towards the primary and the released gravitational potential energy is absorbed by the envelope, which is subsequently ejected \citep[][]{taam2000, Han+2002, Han+2003}. The final geometrical configuration between the components depends on the initial mass ratio and separation. These objects are rare, only a few dozen systems are known so far \citep[see e.g.][]{Almeida+2012, Ostensen+2013, Schaffenroth+2014, Heber2016PASP}. As they provide a direct way to measure the geometrical and physical properties of the components, detailed studies of such systems are crucial to test the theories on formation and evolution of these kind of systems. For a recent review on HW Vir systems, see \citet[][]{Heber2016PASP}. 

The primary star in HW Vir can be classified in three types: hot subdwarf B (hereafter sdB) star, low-mass white dwarf (LMWD) and extremely low-mass (ELM) white dwarf. It depends essentially in how much mass is stripped from the primary star during the common envelope phase and how massive is the remaining core. While most the sdB are core helium-burning stars with a canonical mass of approximately 0.47\,M$_{\odot}$, the LMWDs and ELM white dwarfs do not have enough mass to ignite the helium (He) in their cores. The LMWD and ELM white dwarf have mass around 0.3\,M$_{\odot}$ and 0.2\,M$_{\odot}$ \citep[][]{Brown+2010, Heber2016PASP}, respectively. The spectra of sdB, LMWD, and ELM white dwarf are very similar: all of them populate the extreme horizontal branch (EHB), in the Hertzsprung-Russell (HR) diagram. Therefore, it is necessary to estimate the primary mass to correctly identify its evolutionary status.

HS\,2231+2441 (2MASS\,J22342148+2456573) is an HW~Vir system with an orbital period of $\sim$0.1 d. This system has been studied in two conference papers by \citet[][]{Ostensen+2007} and \citet[][]{Ostensen+2008}. In the first study, the authors showed two possible solutions for this system using one photometric band and 27 epochs of spectroscopic data. They concluded that the most plausible solution provides a secondary mass of 0.075~M$_{\odot}$, when considering a canonical mass 0.470~M$_{\odot}$ for the primary star. In the second one, \citet[][]{Ostensen+2008} presented a new solution by analysing the same data set and obtained masses less than 0.3~M$_{\odot}$ and 0.10~M$_{\odot}$ for the primary and secondary, respectively.

In this study, we improve the solution for HS\,2231+2441 by using BVR$_C$I$_C$ photometry and 44 epochs of phase-resolved optical spectroscopy. This paper is organised as follows: In Section~\ref{sec:ob_data}, we describe the observations and data reduction. The analysis and results of the spectroscopic and photometric data are shown in Section~\ref{sec:ana_resul}. Section~\ref{sec:dis} presents a discussion about the masses and evolution status of the system. Finally, in Section~\ref{sec:con}, we summarise our results.

\section{Observations and Data Reduction}\label{sec:ob_data}

\subsection{Photometry}

Photometric data of HS\,2231+2441 in the B, V, R$_C$, and I$_C$ bands were collected between August, 2010 and August, 2013 using the facilities of Observat\'orio do Pico dos Dias (OPD/LNA), in Brazil. The observations were performed using an Ikon-L CCD camera attached to the 1.6-m and 0.6-m telescopes. To remove systematic effects from the CCD data, we collected typically $\sim$30 bias frames and $\sim$30 dome flat-field images at each night of observation. These observations are described in Table \ref{table:1}. 

\begin{figure}
 \centering
 \resizebox{\hsize}{!}{\includegraphics[angle=-90]{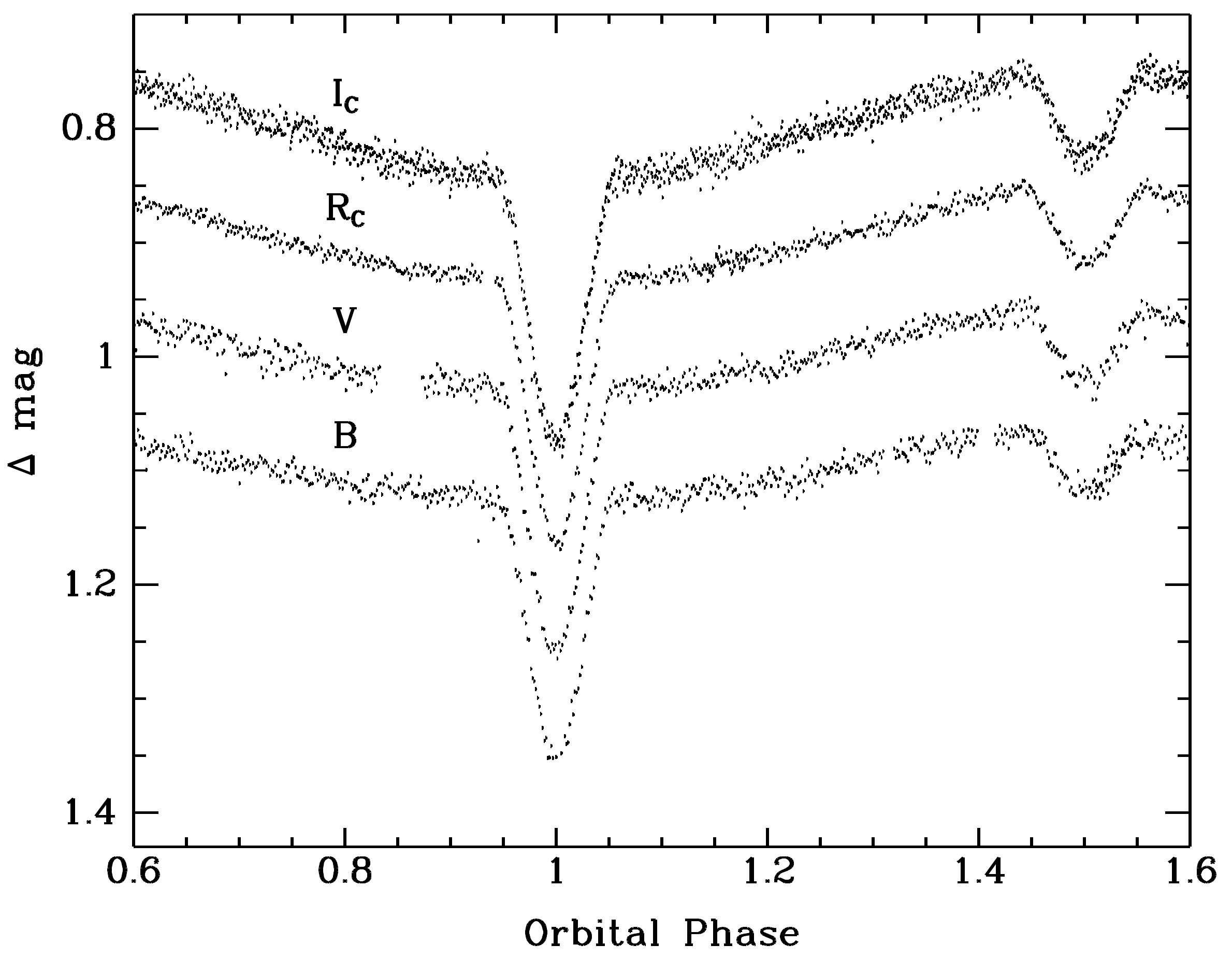}}
 \caption{Normalised light curves of HS\,2231+2441 in the B, V, R$_C$, and I$_C$ bands 
          folded on the 0.1105879 days orbital period. The light curves have arbitrary displacements in $\Delta$\,mag for better visualisation.}
 \label{light_curve}
\end{figure}

Data reduction was done by using usual \verb"IRAF"\footnote{IRAF is distributed by the National Optical Astronomy Observatories, which are operated by the Association of Universities for Research in Astronomy, Inc., under cooperative agreement with the National Science Foundation.} tasks. We built an automated computational procedure that subtracts a master median bias image from each program image and divides the result by a normalised flat-field frame. As the HS\,2231+2441 field is not crowded, aperture differential photometry is a suitable technique to obtain the relative magnitude between our target and a set of constant stars in the field. The result of this procedure can be seen in a sample of normalised light curves (see Figure \ref{light_curve}), folded on the orbital period of HS\,2231+2441, which is derived in Section~\ref{ephemeris}.  

\begin{table}
\caption{Log of the photometric observations of HS\,2231+2441.}  
\label{table:1} 
\centering                   
\begin{tabular}{l c c c c}   
\hline\hline                 
Date~~~~~ & $N$ & \ t$_{\rm exp}$(s) & Telescope & Filter  \\   
\hline                        
  2010 Aug 19  & 300 & 30 & 0.6-m & R$_C$ \\
  2010 Aug 20  & 295 & 30 & 0.6-m & V \\
  2011 Jul 07  & 650 & 10 & 1.6-m & R$_C$ \\
  2011 Jul 08  & 840 & 10 & 1.6-m & I$_C$ \\
  2011 Aug 15  & 900 & 5  & 1.6-m & B \\
  2012 Aug 08  & 600 & 10 & 1.6-m & V \\
  2012 Aug 09  & 630 & 10 & 1.6-m & B \\
  2013 Jul 29  &  644 & 5 & 1.6-m & I$_C$ \\ 
  2013 Aug 01  & 1230 & 4 & 1.6-m & I$_C$ \\
\hline                                   
\end{tabular}
\end{table}

\subsection{Spectroscopy}
Raw spectroscopic data of HS\,2231+2441 were retrieved from the Isaac Newton Group public archive\footnote{http://casu.ast.cam.ac.uk/casuadc/ingarch/query}. The spectroscopic observations were performed with the ISIS spectrograph attached to the 4.2-m William Herschel Telescope at Roque de los Muchachos Observatory. Forty four spectra were collected using the 600 l/mm grating and integration times of 5 and 10 minutes. The spectral coverage is 3600-5000 \AA, with 0.45 \AA~FWHM~resolution. Bias and flat-field images were collected each night to remove systematic signatures from the CCD detector. Wavelength calibration was done by using Cu-Ar comparison lamp obtained at each night of observation. A summary of the observations is shown in Table~\ref{table:RV}.
 
\begin{figure}
 \centering
 \resizebox{\hsize}{!}{\includegraphics[angle=-90]{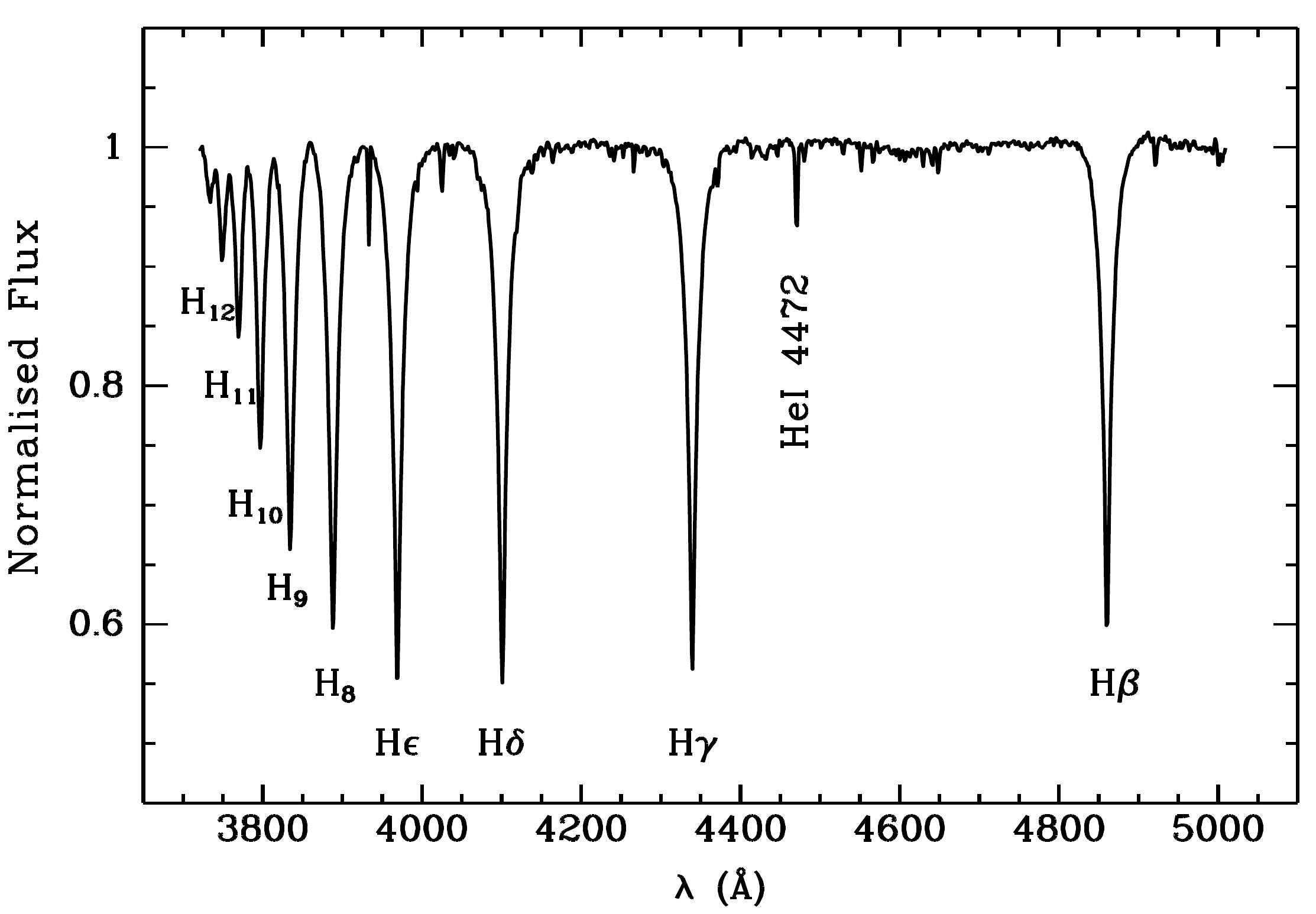}}
 \caption{Average of 44 spectra of HS\,2231+2441 after correcting for orbital motion.}
 \label{fig:spectrum}
\end{figure}

The spectroscopic data reduction was done in the standard way using {\tt IRAF} routines. The procedure includes steps of bias subtraction, normalised flat-field division, optimal extraction, and wavelength calibration. The average of all spectra after Doppler shifting according to the radial velocity orbital solution, see Section \ref{sec:rvsolution}, is shown in Figure~\ref{fig:spectrum}.

\section{Analysis and Results}\label{sec:ana_resul}

\subsection{Differential Photometry}

\subsubsection{Light curves}

The light curves of HS\,2231+2441 (Figure~\ref{light_curve}) are typical of HW~Vir systems, i.e., show a reflection effect and both primary and secondary eclipses. While the secondary eclipse depths increase towards longer wavelength, from $\sim$0.025\,mag in the B-band to $\sim$0.04\,mag in the I$_C$-band, the primary ones do not change significantly with wavelength and have $\sim$0.2\,mag in all bands.

\subsubsection{Ephemeris}\label{ephemeris}

The ephemeris for HS\,2231+2441 was derived by following the same procedure presented in \citet[][]{Almeida+2013}. In short, we measure the mid-eclipse times by modelling the primary eclipse with the Wilson-Devinney code \cite[version 2013;][]{wilson+2010}, which was implemented with a Markov chain with Monte Carlo (MCMC) procedure to obtain the uncertainties. To be consistent, our best solution for the geometrical and physical parameters of HS\,2231+2441 derived in Section~\ref{lcrv} was used as input to the Wilson-Devinney code. The eclipsing times are shown in Table~\ref{timing}.

To fit our mid-eclipse times, we use the linear ephemeris expression 
\begin{equation}
T_{\rm min} = T_0 + E \times P_{\rm orb}, 
\end{equation}
where $T_{\rm min}$ is the predicted times of primary minimum, $T_0$ is a fiducial epoch, $E$ is the cycle count from $T_0$, and $P_{\rm orb}$ is the orbital period. The best result is
\begin{equation}
T_{\rm min}~ (\rm \bf BJD) = 2455428.76185(10) + 0.1105879(1)  \times E.
\label{eq:ephem1}
\end{equation}
The residuals with respect to this ephemeris show no evidence of systematic variations.

\subsection{Spectroscopic analysis}

\subsubsection{Radial velocity solution}
\label{sec:rvsolution}

\begin{figure}
 \resizebox{\hsize}{!}{\includegraphics[angle=-90]{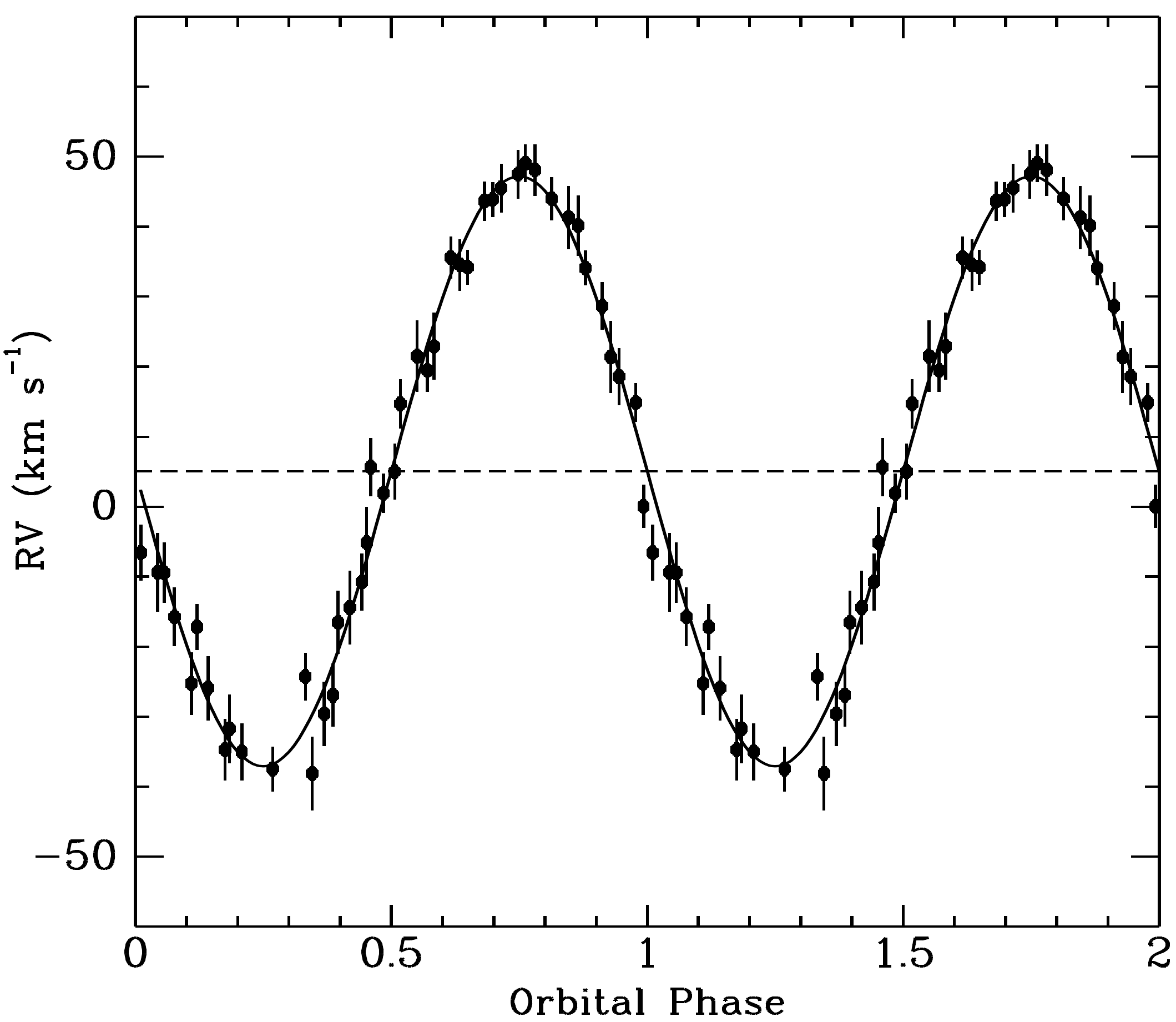}}
 \caption{Radial velocities of the prominent lines in the spectra of HS\,2231+2441 folded on the orbital period. The phases are calculated according to the ephemeris in Equation~\ref{eq:ephem1}. Solid and dashed lines represent the best radial velocity solution and systemic velocity, respectively.}
 \label{fig:RV}
\end{figure}

HS\,2231+2441 spectra only show absorption lines of the primary star (see Figure~\ref{fig:spectrum}). Aiming to obtain the radial velocity amplitude of the primary star, we used the H${\beta}$, H${\gamma}$, H${\delta}$, H${\epsilon}$, H$_{\rm 8}$, H$_{\rm 9}$, H$_{\rm 10}$, and H$_{\rm 11}$ lines to measure the Doppler shifts in the sample of 44 spectra (see Table~\ref{table:RV}). To do that, we used the task {\tt FXCOR} in {\tt IRAF}. Initially, we selected the spectrum that has orbital phase $\phi = 0.01$ as template for the cross-correlation with all others spectra. The resulting radial velocity solution was used to Doppler-shift all individual spectra to the orbital rest frame. Then, all Doppler-shifted spectra were combined to create a new template with better signal-to-noise. This procedure was done a few times, typically 10 runs, until the radial velocity solution converged. Table \ref{table:RV} lists individual radial velocities and Figure \ref{fig:RV} shows the radial velocity curve folded on the orbital phase together with the best solution for a circular orbit. The modelling provides $42\pm1$\,km\,s$^{-1}$ and $5\pm1$\,km\,s$^{-1}$ for the radial velocity semi-amplitude ($K_1$) and systemic velocity ($\gamma$), respectively. Our measurement of $K_1$ is about 14\,\% smaller than the result obtained by \citet[][]{Ostensen+2007} and \citet[][]{Ostensen+2008}, which is $K_1=49.1\pm3.2$ km\,s$^{-1}$. 

\begin{table}
\begin{center}
\begin{small}
\caption{Spectroscopic observations and radial velocity measurements of the primary star.}             
\label{table:RV} 
\begin{tabular}{r r r r r}  
\hline\hline                
UT Date& HJD        & t$_{\rm exp}$ & V~~~~~~~   & Orbital \\   
       & (2450000+) &    (s)        & (km\,s$^{-1}$)~~~ & Phase \\
\hline                        
   2005 Sep 18 & 3632.4102 & 300 & -38.1$\pm$5.3  & 0.35  \\
   2005 Sep 18 & 3632.4147 & 300 & -26.9$\pm$4.5  & 0.39  \\
   2005 Sep 18 & 3632.4183 & 300 & -14.4$\pm$5.3  & 0.42  \\
   2005 Sep 18 & 3632.4220 & 300 &  -5.2$\pm$5.1  & 0.45  \\
   2005 Sep 18 & 3632.4256 & 300 &   1.9$\pm$2.8  & 0.48  \\
   2005 Sep 18 & 3632.4292 & 300 &  14.7$\pm$3.5  & 0.52  \\
   2005 Sep 18 & 3632.4329 & 300 &  21.5$\pm$5.1  & 0.55  \\
   2005 Sep 18 & 3632.4365 & 300 &  22.9$\pm$4.8  & 0.58  \\
   2005 Sep 18 & 3632.4401 & 300 &  35.6$\pm$3.0  & 0.62  \\
   2005 Sep 18 & 3632.4438 & 300 &  34.2$\pm$2.5  & 0.65  \\
   2005 Sep 18 & 3632.4474 & 300 &  43.7$\pm$2.8  & 0.68  \\
   2005 Sep 18 & 3632.4511 & 300 &  45.5$\pm$3.5  & 0.71  \\
   2005 Sep 18 & 3632.4547 & 300 &  47.5$\pm$3.5  & 0.75  \\  
   2005 Sep 18 & 3632.4583 & 300 &  48.1$\pm$3.7  & 0.78  \\  
   2005 Sep 18 & 3632.4620 & 300 &  44.0$\pm$3.1  & 0.81  \\  
   2005 Sep 18 & 3632.4656 & 300 &  41.3$\pm$4.5  & 0.85  \\   
   2005 Sep 18 & 3632.4692 & 300 &  34.1$\pm$2.5  & 0.88  \\   
   2005 Sep 18 & 3632.4729 & 300 &  28.7$\pm$3.4  & 0.91  \\   
   2005 Sep 18 & 3632.4764 & 300 &  18.6$\pm$4.1  & 0.94  \\   
   2005 Sep 18 & 3632.4801 & 300 &  14.9$\pm$2.8  & 0.98  \\   
   2005 Sep 18 & 3632.4838 & 300 &  -6.6$\pm$4.0  & 0.01  \\   
   2005 Sep 18 & 3632.4874 & 300 &  -9.4$\pm$5.6  & 0.04  \\   
   2005 Sep 18 & 3632.4910 & 300 & -15.8$\pm$4.2  & 0.08  \\   
   2005 Sep 18 & 3632.4947 & 300 & -25.3$\pm$4.5  & 0.11  \\ 
   2005 Sep 18 & 3632.4983 & 300 & -25.9$\pm$4.6  & 0.14  \\
   2005 Sep 18 & 3632.5019 & 300 & -34.7$\pm$4.4  & 0.17  \\
   2005 Sep 18 & 3632.5056 & 300 & -35.0$\pm$4.1  & 0.21  \\   
   2010 Aug 29 & 5437.5391 & 600 & -29.6$\pm$4.6  & 0.37  \\ 
   2010 Aug 29 & 5437.5473 & 600 & -10.8$\pm$4.1  & 0.44  \\   
   2010 Aug 29 & 5437.5543 & 600 &   5.0$\pm$4.0  & 0.51 \\  
   2010 Aug 29 & 5437.5614 & 600 &  19.5$\pm$3.1  & 0.57 \\  
   2010 Aug 29 & 5437.5684 & 600 &  34.5$\pm$3.7  & 0.63 \\
   2010 Aug 29 & 5437.5755 & 600 &  43.9$\pm$2.5  & 0.70 \\
   2010 Aug 29 & 5437.5826 & 600 &  49.1$\pm$2.7  & 0.76 \\   
   2010 Aug 29 & 5437.5940 & 600 &  40.2$\pm$4.3  & 0.86 \\   
   2010 Aug 29 & 5437.6010 & 600 &  21.4$\pm$5.2  & 0.93 \\
   2010 Aug 29 & 5437.6081 & 600 &   0.4$\pm$3.1  & 0.99 \\   
   2010 Aug 29 & 5437.6151 & 600 &  -9.4$\pm$4.3  & 0.06 \\   
   2010 Aug 29 & 5437.6222 & 600 & -17.2$\pm$3.3  & 0.12 \\   
   2010 Aug 29 & 5437.6292 & 600 & -31.7$\pm$4.9  & 0.18 \\   
   2010 Aug 29 & 5437.6386 & 600 & -37.5$\pm$3.2  & 0.27 \\   
   2010 Aug 29 & 5437.6456 & 600 & -24.3$\pm$3.4  & 0.33 \\   
   2010 Aug 29 & 5437.6526 & 600 & -16.5$\pm$4.5  & 0.40 \\   
   2010 Aug 29 & 5437.6597 & 600 &   5.6$\pm$4.2  & 0.46 \\   
\hline                              
\end{tabular}
\end{small}
\end{center}
\end{table}

\subsubsection{Spectroscopic parameters}\label{spectra_fitting}

In order to obtain the atmospheric parameters of the HS\,2231+2441 primary star, we compared the observed spectrum (see Figure~\ref{fig:spectrum}) with a grid of synthetic spectra retrieved from the web-page of TheoSSA\footnote{http://dc.g-vo.org/theossa}. The synthetic spectra were generated from non-local thermodynamic equilibrium (NLTE) models with zero metallicity. The grid comprises 52 values of effective temperatures in the range 25000 -- 35000\,K with 500\,K steps, 16 surface gravities, 5.2 $\leq$ $\log g$ $\leq$ 6.0 with 0.05 dex steps, 10 He abundances, 0.001 $\leq$ $n(He)/n(H)$ $\leq$ 0.01, with 0.0005 dex steps, and 20 projected rotational velocities, 0 $\leq v_{\rm rot} \leq 200\,\rm km\,s^{-1}$, with 10\,km\,s$^{-1}$steps. 

\begin{figure}
 \centering
 \resizebox{\hsize}{!}{\includegraphics[angle=-90]{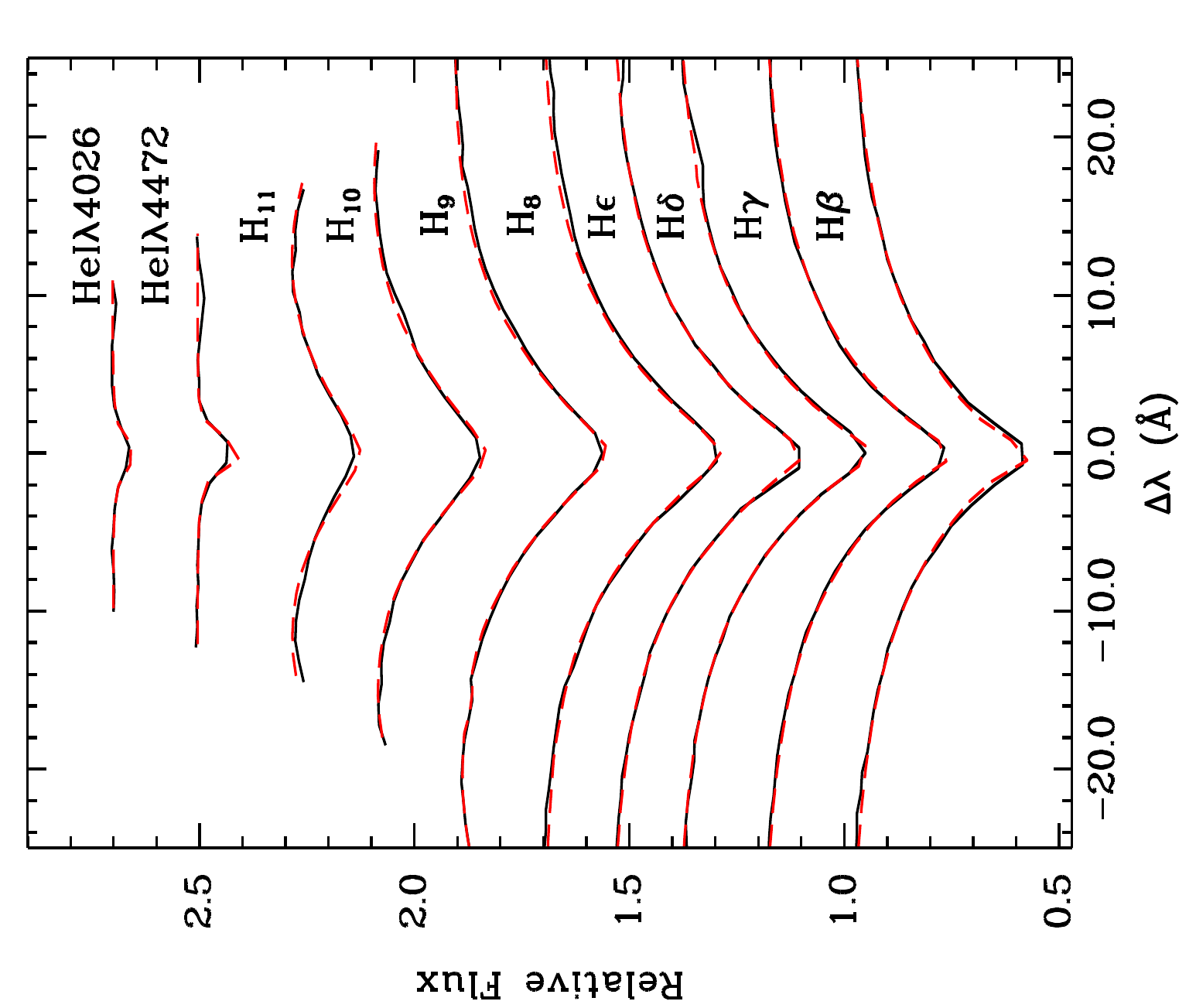}}
 \caption{The best fit to the Balmer and helium lines used to derive effective temperature, surface gravity, helium abundance, and rotational velocity. The observed spectrum lines are presented with solid line and the dashed line represents the best synthetic spectrum.}
\label{fig:fitspectrum}
\end{figure}

To choose the synthetic spectrum that best matches the observed one, we convolved all synthetic spectra with the FWHM (0.5\AA) of the ins\-trumental profile obtained from the spectroscopic observation. The spectral regions around the Balmer (H$\beta$ to H$_{11}$) and helium (HeI${\lambda 4026}$, HeI${\lambda4471}$) lines were used in the fitting procedure to determine the effective temperature, surface gravity, He abundance, and rotational projected velocity. A minimisation using the $\chi^2$ technique was done in the last step. The best fit yields $T_{\rm eff} = 28500\pm500$\,K, $\log g = 5.40\pm0.05$, $\log(n(He)/n(H)) = -2.52\pm0.07$, and $v_{\rm rot} = 70\pm10\,\rm km\,s^{-1}$. The adopted error bar in each parameter is the step in its respective grid. With the exception of n(He)/n(H), our results are in agreement with those obtained by \citet[][]{Ostensen+2007,Ostensen+2008}. The projected velocity is consistent with a synchronised system, i.e. $P_{\rm orb} = P_{\rm rot} = 2\pi R / v_{\rm rot}$ (see Table~\ref{sytem:parameters}). Figure~\ref{fig:fitspectrum} shows the observed lines superimposed with the best synthetic spectrum.

\begin{table}
\caption{Eclipse timings of HS\,2231+2441.} 
\label{timing}      
\centering          
\begin{tabular}{l c c }
\hline\hline           
Cycle &Eclipse timing    & O-C                           \\    
      &JD(TDB) 2450000+  &  (s)                            \\
\hline                       
0     &  5428.76188(3)   &  2.6                        \\  
9     &  5429.75719(6)   &  4.3                        \\  
2912  &  5750.79366(2)   & -1.5                        \\
2921  &  5751.78896(2)   & -0.7                        \\
3264  &  5789.72060(2)   &  0.0                        \\
6510  &  6148.688830(9)  &  5.6                        \\
6519  &  6149.684087(9)  & -0.3                        \\
9748  &  6506.772247(9)  &  0.5                        \\
\hline                                 
\end{tabular}
\end{table}

\subsection{Light and radial velocity curves analysis}\label{lcrv}

Simultaneous modelling of multi-band light curves and radial velocity curves is a powerful tool to derive geometrical and physical parameters of eclipsing binaries \citep[see e.g.][]{Wilson1979, Almeida+2012,Almeida+2015}. We fit the light curves in the B, V, R$_C$, and I$_C$ bands and the primary star radial velocity curve of the HS\,2231+2441 binary using the latest version of the Wilson Devinney code \citep[WDC;][]{wilson+2010}. The WDC is a generator of synthetic multi-band light curves and radial velocity curves of binary systems. To optimise the fitting procedure, we incorporate the genetic algorithm {\tt PIKAIA} \citep[][]{Charbonneau1995} to the WDC to search for a global solution and a MCMC procedure to sample around their expected values and obtain their uncertainties.

The main parameters adjustable by the WDC are: orbital period ($P_{\rm orb}$), epoch ($T_0$), mass ratio ($q=M_2/M_1$), inclination ($i$), adimensional potentials and temperature of both components ($\Omega_1$, $\Omega_2$, $T_1$, and $T_2$). However, the WDC allows one to fit about 60 parameters, which refer to physical and geometrical properties of the binary and also to a possible third component. Therefore, one needs to constrain them as much as possible with prior information from photometry, spectroscopy and theory. In our case, the orbital period and epoch are already known from the light curves (see Section~\ref{ephemeris}). From the spectroscopic analysis, we derive the effective temperature of the primary star (see Section~\ref{spectra_fitting}), and together with theoretical information, we can constrain $q$ using $K_1$ and the mass function,
\begin{equation}
 f(m_2) = \frac{M_1(q\sin\,i)^3}{(1+q)^2} = \frac{(1-e^2)}{2\pi G}K_1^3P_{\rm orb},
  \label{eq:mass_function}
\end{equation}
where $G$ is the gravitational constant. We adopted the ranges $0.1\,M_{\odot}<M_1<0.8\,M_{\odot}$ -- which covers the mass range for sdBs, LMWDs, and ELM white dwarfs \citep[][]{Heber2016PASP} -- and $75^{\circ}<i<90^{\circ}$ to search for the primary mass and orbital inclination, respectively. Assuming a circular orbit ($e=0$), the estimated mass ratio range is 0.11 -- 0.25.   

Although the orbital period is very short, the shape of its light curve shows a detached configuration. Therefore, the WDC mode 2 is the most suitable because it does not constrain the Roche configuration. The luminosity of the secondary component was computed by using a stellar atmosphere radiation. Gravity darkening exponent ($\beta_1$) and the bolometric albedo ($A_1$) of the primary star were set to 1, while the gravity darkening exponents of the secondary component ($\beta_2$) was fixed to 0.3 \citep[][]{Rafert+1980}. The linear limb darkening coefficients $(\alpha)$ from \citet[][]{Claret+2011}, which assume different values for the different photometric bands (see Table~\ref{sytem:parameters}), were linearly extrapolated for both stars. As pointed out in previous studies \citep[see e.g.][]{Almeida+2012,For+2010}, the albedo of the secondary star in HW Vir systems can assume nonphysical values, i.e. $A_2 > 1$. This happens because the reflected-reradiated spectral energy rate from the secondary star increases towards longer wavelengths and may reach values greater than 1. To take this effect into account, we adopted the secondary albedo in all bands as a free parameter.

\begin{figure}
 \resizebox{\hsize}{!}{\includegraphics[angle=-90]{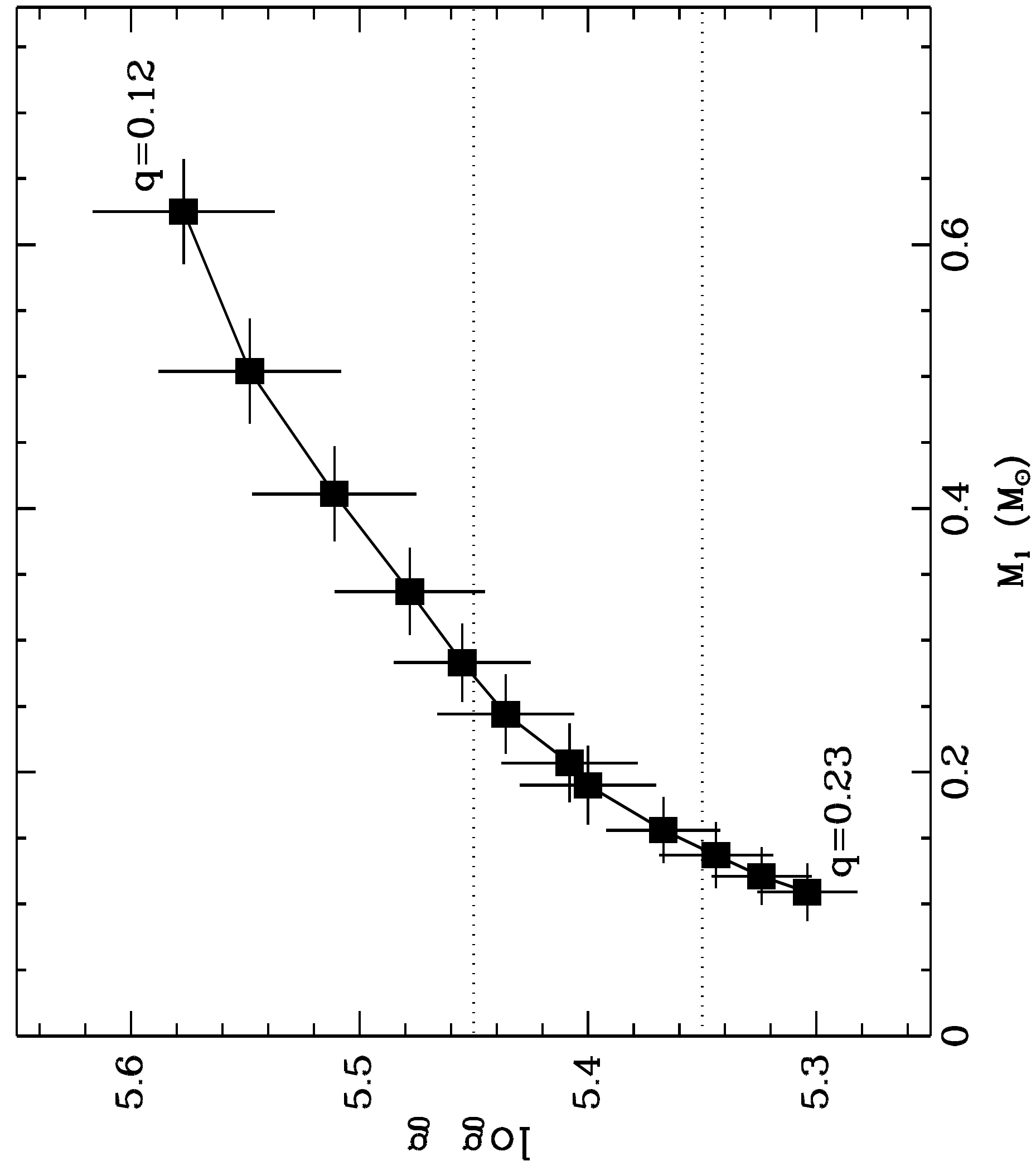}}
 \caption{WDC solutions for surface gravity and mass of the primary star using fixed values of $q$ are represented by squares. Horizontal dotted lines represent the lower and upper limits to surface gravity derived from the spectroscopic modelling. The photometric and spectroscopic surface gravities are equal for $q=0.19$.}
 \label{fig05}
\end{figure}

\begin{figure}
 \resizebox{\hsize}{!}{\includegraphics[angle=-90]{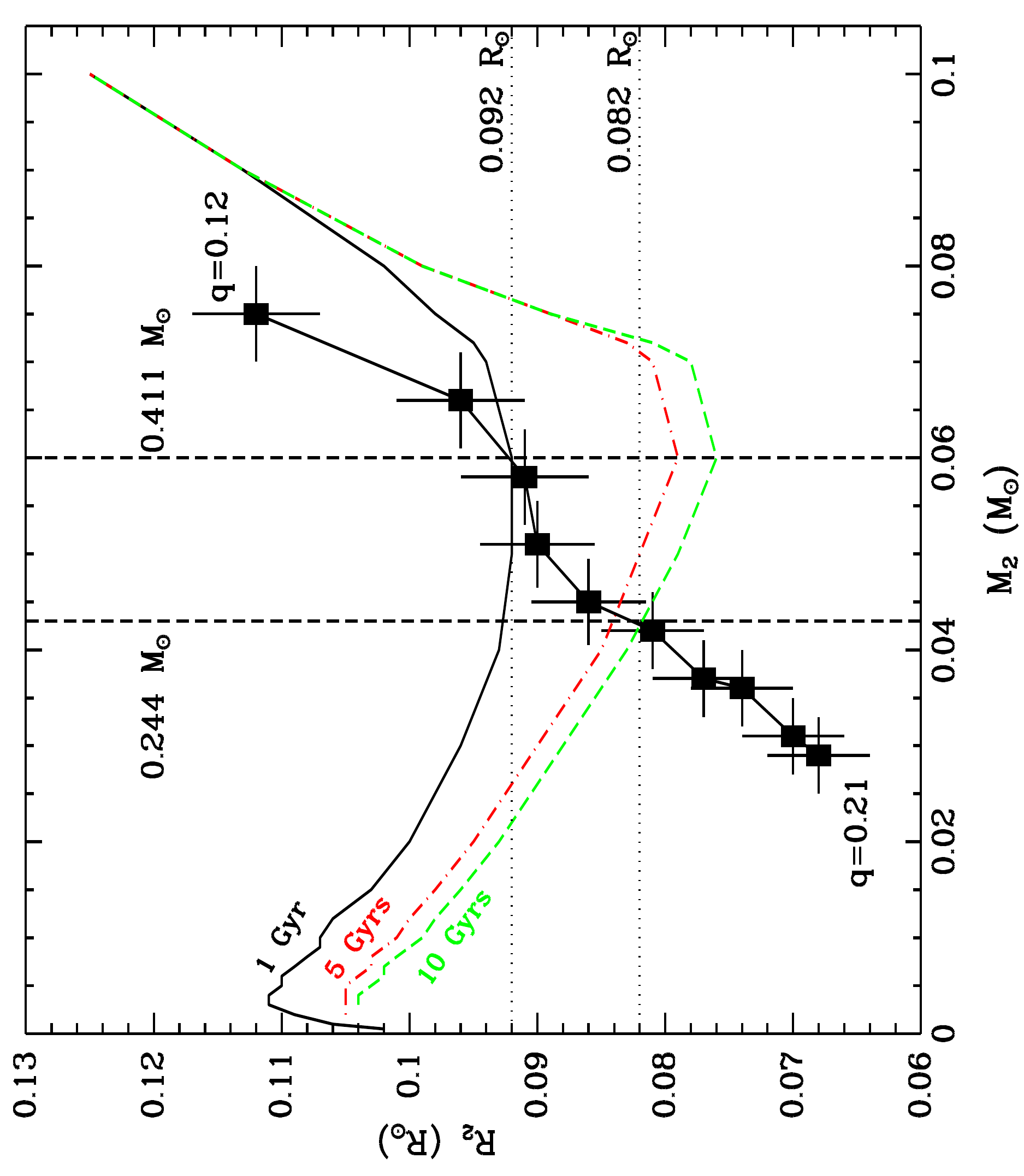}}
 \caption{ Mass-radius relationship for low-mass and brown dwarf stars. WDC solutions for the secondary mass and radius using fixed values of $q$ are shown using squares. Theoretical mass-radius relationships for 1, 5, and 10 Gyrs calculated by \citet[][]{Baraffe+2003} are shown with solid, dash-dotted and dashed lines, respectively. Horizontal dotted and vertical dashed lines represent the lower and upper limits for the secondary radius and the primary mass considering the possible theoretical range for the mass and radius of the secondary star, respectively.}
 \label{fig06}
\end{figure}

Finally, the remaining adjustable parameters for HS\,2231+2441 are: mass ratio, orbital inclination, separation between the components ($a$), the adimensional potentials and effective temperature of the secondary. However, as the mass ratio is highly correlated with the other parameters \citep[see e.g.][]{Schaffenroth+2014,Schaffenroth+2015}, we performed a grid of solutions keeping $q$ fixed. The grid has steps of 0.01 in mass ratio covering the range pointed out above. The models for different $q$ have very similar chi-square values and the choice of the most probable solutions was done as described below. Figure~\ref{fig05} compares the surface gravity of the WDC fittings (photometric solution, from hereafter) with that obtained from the spectral lines profiles (spectroscopic solution, for short). The photometric solution coincides with the spectroscopic one for $q = 0.19$. A different approach is presented in Figure~\ref{fig06}, which shows the secondary mass and radius from the photometric solutions and the theoretical models for low-mass stars and brown dwarfs from \citet[][]{Baraffe+2003}. The solutions are consistent with the models for $0.14 < q < 0.17$. Considering that the age for compact hot stars with progenitor less massive than 1.2 $M_{\odot}$ is larger than 5 Gyrs \citep[][]{Schindler+2015}, the most plausible solution is obtained with $q = 0.16$. Therefore, we list the fitted and derived parameters for these two solutions in Table~\ref{sytem:parameters}.

\begin{figure}
 \resizebox{\hsize}{!}{\includegraphics[angle=0]{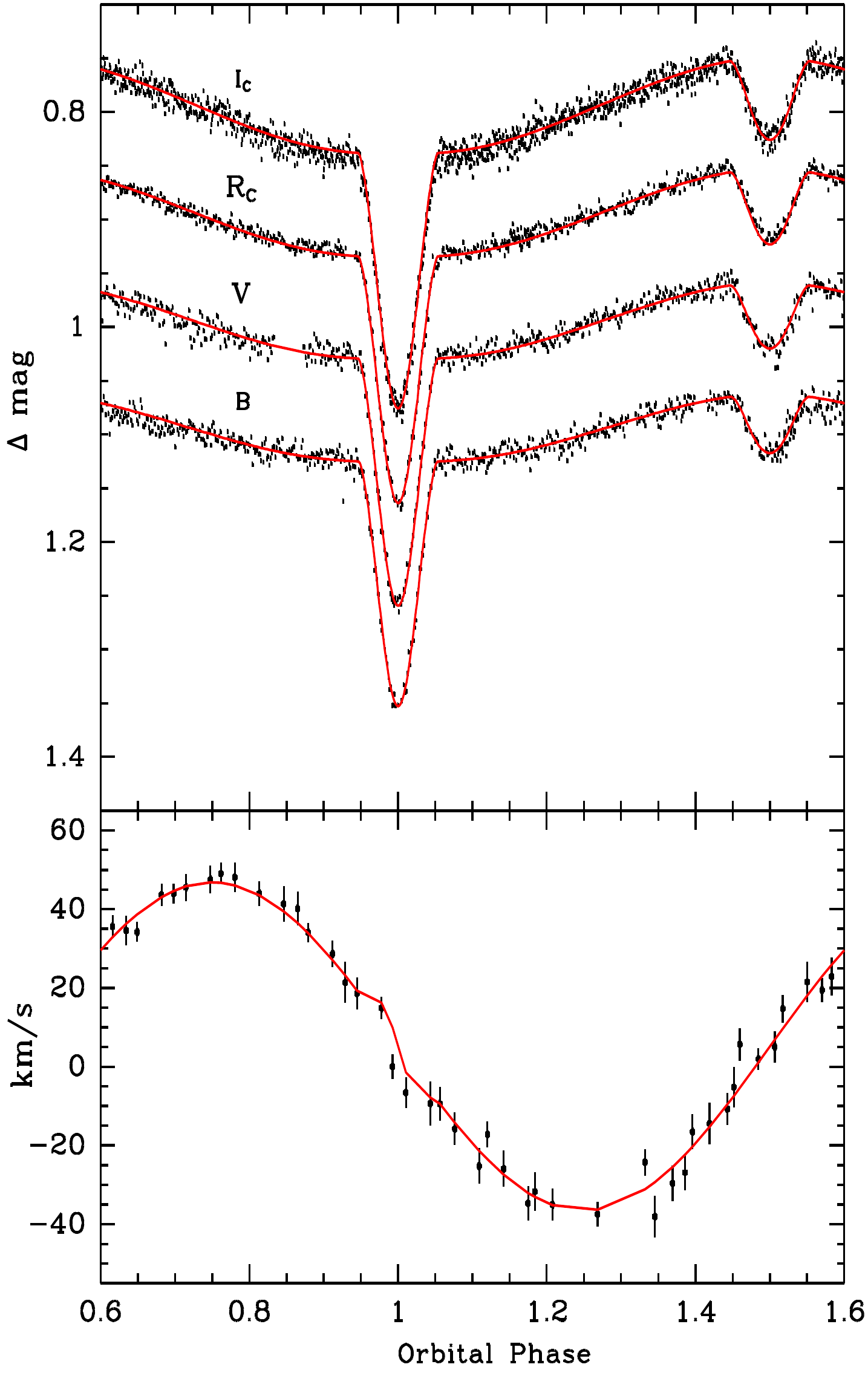}}
 \caption{The best simultaneous fit to the light curves in the B, V, R$_C$, and I$_C$ bands and primary radial velocity curve performed with the Wilson-Devinney code.}
 \label{lc_rv}
\end{figure}

\begin{figure}
 \resizebox{\hsize}{!}{\includegraphics{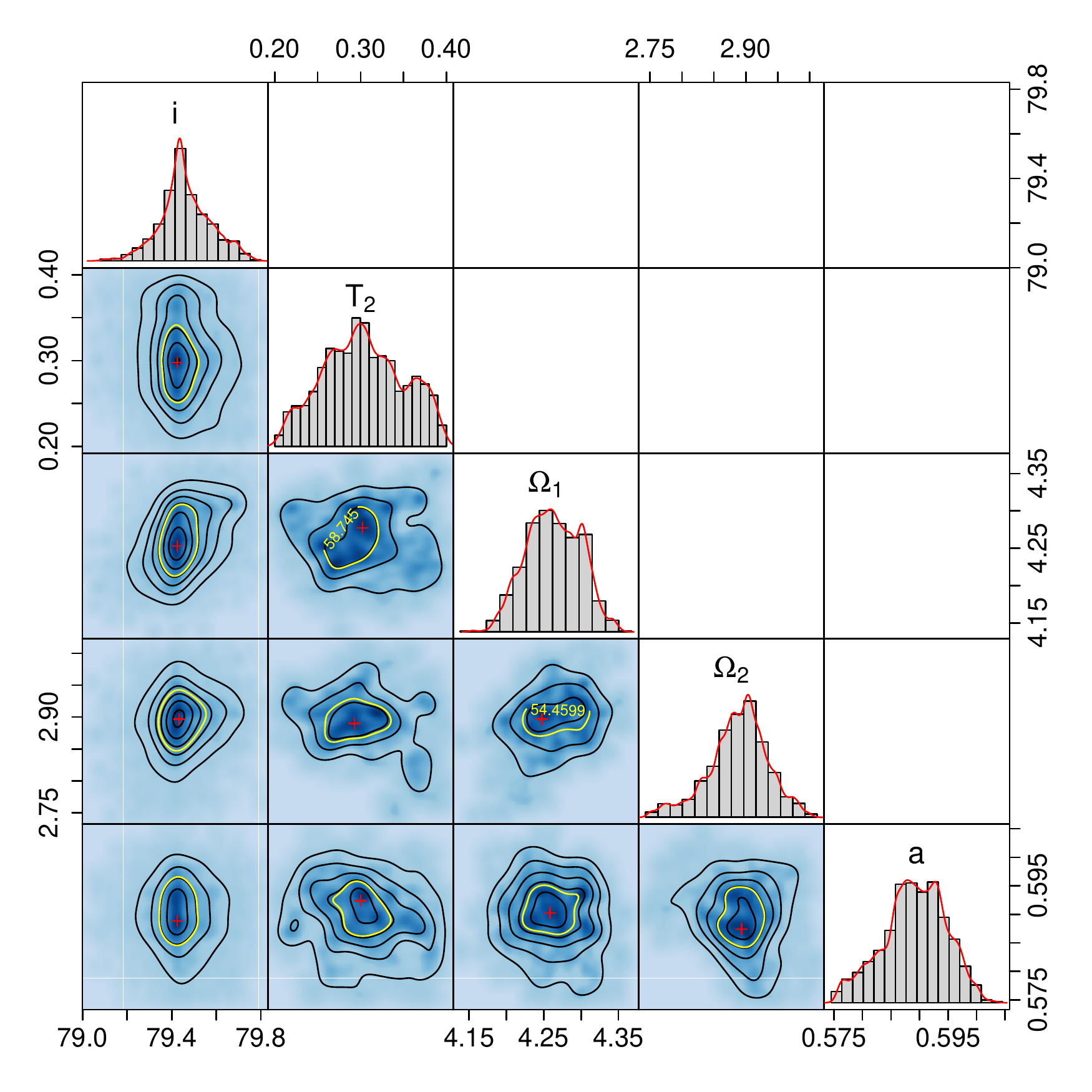}}
 \caption{
 Joint distributions of the {\it a posteriori} probability densities for the main parameters fitted in the simultaneously analysis of the HS\,2231+2441 light and radial curves. Our MCMC has $2\times10^4$ iterations sampling the regions around the best solution obtained with $q=0.19$ found by the PIKAIA algorithm.}
 \label{histo}
 \end{figure}

Figure~\ref{lc_rv} shows the best fit for $q=0.19$ together with the light curves in the B, V, R$_C$, and I$_C$ bands and the primary radial velocity curve. Both solutions have indistinguishable fits and $\chi^2_{\rm red} = 1.03$. Figure~\ref{lc_rv} (bottom panel) shows a feature visible between the phases 0.95 and 1.05, similar to a sinusoidal curve superimposed to the radial velocity curve. It is known as the Rossiter-McLaughlin effect and occurs when the secondary passes in front of a rotating primary and vice versa, if both stars are visible \citep[][]{Ohta+2005}. In our fitting, we take this effect into account adopting the simplified case, i.e. circular orbit and synchronised and corotational components. As an example, Figure~\ref{histo} shows the {\it a posteriori} probability densities of the main fitted parameters for the solution obtained with $q=0.19$. 

\begin{table}
\small
\begin{center}
\caption{System parameters of the best model fit to photometric light curves in the B, V, R$_C$, I$_C$ bands and the primary star radial velocity curve of HS\,2231+2441.}         
\label{sytem:parameters}    
\begin{tabular}{l c r}    
\hline
\hline                 
 Parameter & Value    &       \\     
\hline
Fixed Parameters    &  Solution 1            &  Solution 2 \\
\hline
q$(M_2/M_1)$         &   0.190           &      0.160  \\
T$_1$(K)             &   28500           &      28500  \\
$\alpha^a_1$(B)      &   0.290           &      0.290  \\   
$\alpha^a_1$(V)      &   0.256           &      0.256  \\   
$\alpha^a_1$(R$_C$)  &   0.222           &      0.222  \\   
$\alpha^a_1$(I$_C$)  &   0.188           &      0.188  \\   
$\alpha^a_2$(B)      &   0.727           &      0.677  \\   
$\alpha^a_2$(V)      &   0.742           &      0.651  \\   
$\alpha^a_2$(R$_C$)  &   0.705           &      0.624  \\   
$\alpha^a_2$(I$_C$)  &   0.594           &      0.527  \\
$\beta^b_1$          &    1              &        1  \\
$\beta^b_2$          &   0.3             &        0.3  \\
$A^c_1$              &    1              &        1  \\
\hline
Adjusted Parameters &  Solution 1            &  Solution 2 \\
\hline 
$\Omega^d_1 $       &         4.31$\pm$0.05  &      4.24$\pm$0.06   \\
$\Omega^d_2$        &         2.91$\pm$0.03  &      2.68$\pm$0.02   \\
T$_2$(K)            &         3010$\pm$460   &      3410$\pm$500   \\
i~($^{\circ}$)      &         79.4$\pm$0.2   &      79.6$\pm$0.1   \\
a$^e$ (R$_{\odot}$) &        0.59$\pm$0.01   &      0.67$\pm$0.02   \\ 
A$^c_2$(B)          &        1.243$\pm$0.07  &      1.39$\pm$0.09   \\
A$^c_2$(V)          &        1.305$\pm$0.06  &      1.28$\pm$0.08   \\
A$^c_2$(R$_C$)      &        1.574$\pm$0.05  &      1.76$\pm$0.06   \\
A$^c_2$(I$_C$)      &        1.797$\pm$0.05  &      1.87$\pm$0.05   \\
\hline   
Derived parameters \\
\hline
M$_1$ (M$_{\odot}$)      &      0.190$\pm$0.006  &     0.288$\pm$0.005   \\
M$_2$ (M$_{\odot}$)      &      0.036$\pm$0.004  &     0.046$\pm$0.004   \\
R$_1$ (R$_{\odot}$)      &      0.144$\pm$0.004  &     0.165$\pm$0.005   \\ 
R$_2$ (R$_{\odot}$)      &      0.074$\pm$0.004  &     0.086$\pm$0.004   \\
log~g$_1\,(\rm cm\,s^{-2})$ &      5.40$\pm$0.03    &     5.46$\pm$0.03 \\
log~g$_2\,(\rm cm\,s^{-2})$ &      5.25$\pm$0.07    &     5.23$\pm$0.006 \\
v$^f_{\rm rot;1}$        &      65.9$\pm$1.9     &      75.5$\pm$2.3  \\
v$^f_{\rm rot;2}$        &      33.9$\pm$1.9     &      39.4$\pm$1.9  \\
\hline
\end{tabular}
\end{center}
$^a$ Linear limb darkening coefficient from \citet[][]{Claret+2011};\\
$^b$ Gravity darkening exponent;\\
$^c$ Bolometric albedo;\\
$^d$ Adimensional potential; \\
$^e$ Components separation;\\
$^f$ Rotational velocity adopting synchronised rotation ($P_{\rm orb} = P_{\rm rot}$).
\end{table}

\section{Discussion}\label{sec:dis}

\subsection{Evolution Status of the sdB star}

According to \citet[][]{Han+2003}, a sdB star in HW Vir systems can be formed in the following channel. The two components of the binary are on the main sequence and the primary is the most massive. When the primary ascends to the red giant branch (RGB) a dynamical mass transfer begins. This leads to the CE and a spiral-in phase. The primary envelope absorbs the released gravitational potential energy and it is subsequently ejected. The sdB star is originated if the core of the giant still burns helium. The final result is a short-period binary with a sdB plus a main sequence companion. 




In the channel pointed out above, the resultant sdB star will have a mass larger than $\sim$0.47~$\rm M_{\odot}$, which is the lower limit to ignite the helium in the its core \citep[][]{Han+2003}. However, a LMWD or an ELM white dwarf star can be formed through the so-called post-Red Giant Branch (post-RGB) scenario \citep[][]{dreibe+1998,Han+2003}. This channel occurs when the remaining mass of the primary after the first CE phase is insufficient to ignite helium. Then, the primary star will evolve through the sdB region and form a helium-core WD.

\begin{figure}
 \resizebox{\hsize}{!}{\includegraphics[angle=-90]{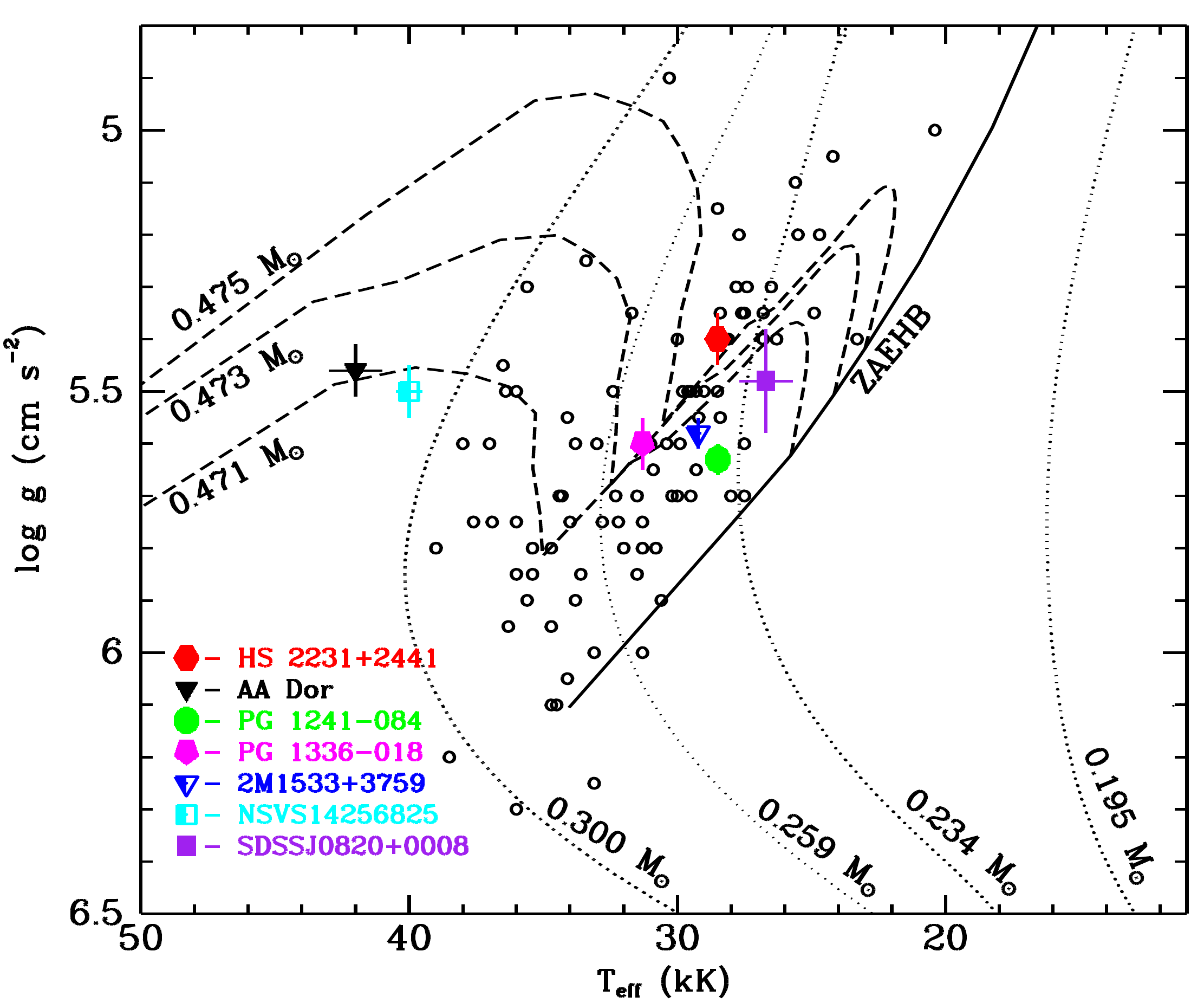}}
 \caption{Location of the HS\,2231+2441 primary star in the T$_{\rm eff}-\log g$ diagram. A few  primary stars in HW~Vir systems are shown with different symbols (see legend). Open circles represent isolated sdB stars collected from \citet[][]{Edelmann2003}. Evolutionary tracks to different masses in the post-EHB evolution \citep[][]{Dorman1993} and post-RGB evolution \citep[][]{dreibe+1998} are shown with dashed and dotted lines, respectively. Solid line represents zero age at extreme horizontal branch (ZAEHB).}
 \label{txlogg}
\end{figure}

In order to check which evolutionary model is the most likely for the HS\,2231+2441 primary star, we use the $\log g$ versus effective temperature ($T_{\rm eff}$) diagram. Figure \ref{txlogg} shows the location of the primary component, some sdB stars in HW Vir systems, and a sample of single sdB stars presented in \citet[][]{Edelmann2003}. Post-EHB and post-RGB evolutionary tracks presented by \citet[][]{Dorman1993} and \citet[][]{dreibe+1998}, respectively, are also shown. Considering the mass derived for the primary star in both solutions (see Table~\ref{sytem:parameters}), the most plausible scenario is the post-RGB evolutionary track placing the HS\,2231+2441 primary star in the LMWD or ELM white dwarf regime. 

\subsection{Masses of the HS\,2231+2441 components}

In recent years, several dozen of these LMWDs and ELM white dwarfs have been reported in the literature \citep[see e.g.][]{Rebassa-Mansergas+2011, Gianninas+2014, Gianninas+2014b}. These objects were found as part of binary systems and their companions can be of several types, e.g. neutron stars, white dwarfs, and main sequence stars \citep[see e.g.][]{Gianninas+2014, Brown+2016, Heber2016PASP}. The first reported ELM white dwarf is the companion of a neutron star in the millisecond pulsar PSR J1012+5307 \citep[][]{vanKerkwijk+1996}. Later, \citet[][]{dreibe+1998} determined the mass of this object $M = 0.19\pm0.02\,\rm M_{\odot}$. \citet[][]{Rebassa-Mansergas+2011} compared the mass distribution of post-common envelope binaries (PCEBs) and wide white dwarf plus main sequence binaries from the Sloan Digital Sky Survey and confirmed that the majority of LMWDs reside in close binary systems. They showed that, despite the fact that the mass distribution of the whole sample displays two peaks near 0.55~M$_{\odot}$ and 0.4 M$_{\odot}$, the mass distribution of the PCEBs has a concentration of systems towards the low-mass side. \citet[][]{Heber2016PASP} suggested that the ELM and low-mass white dwarfs have mass around 0.2 and 0.3~M$_{\odot}$, respectively. Therefore, both solutions for HS\,2231+2441, which provide primary mass of 0.19~M$_{\odot}$ and 0.288~M$_{\odot}$ (see Table~\ref{sytem:parameters}), place the primary component as either an ELM white dwarf or a LMWD.

The confirmed dM-star/brown-dwarf companions in HW Vir systems have a mass distribution  concentrated around 0.1\,M$_{\odot}$ \citep[][]{Kupfer+2015}. In the lower part of the mass distribution, there are a few cases with confirmed brown dwarfs: SDSS\,J162256$+$473051, SDSS\,J082053.53+000843.4, and V2008-1753, which have secondary masses 0.060M$_{\odot}$ \citep[][]{Schaffenroth+2014}, 0.068M$_{\odot}$ \citep[][]{Geier+2011}, and 0.047\,M$_{\odot}$ \citep[][]{Schaffenroth+2015}, respectively. Thus, the secondary star of HS\,2231+2441, considering both solutions, has one of the lowest masses, $\sim$0.036 and $\sim$0.046\,M$_{\odot}$, for a companion object in HW Vir systems known so far.

If we assume the sdB canonical mass ($0.47~\rm M_{\odot}$) for the HS\,2231+2441 primary, the mass ratio would be approximately equal to 0.135. Concerning the primary, its surface gravity would be $\log g \sim 5.53$, which is in disagreement with the measured spectroscopic surface gravity (Figure~\ref{fig05}). Moreover, the radius of the secondary star would be much larger ($\sim$0.094\,R$_{\odot}$) than the expected from theoretical mass-radius relationship (see Figure~\ref{fig06}). The mass of the secondary component would be 0.063\,M$_{\odot}$, still consistent with a brown dwarf.

\section{Summary}\label{sec:con}

In this study, we characterise the HW~Vir system HS\,2231+2441 using photometric and spectroscopic data. This system has a short orbital period, $P_{\rm orb}=0.11$~d, and shows a typical HW Vir light curve, i.e. both primary and secondary eclipses and reflection effect. Spectroscopic analysis enabled us to derive the following properties of the HS\,2231+2441 primary: the semi-amplitude of the radial velocity curve, $42\pm1.0$ km/s, the effective temperature, T$_{\rm eff} = 28500\pm500$ K, the surface gravity, $\log g = 5.40\pm 0.05$, and the helium abundance, $\log(n(He)/n(H)) = -2.52\pm0.07$.  Furthermore, we derive the projected rotational velocity, $v_{\rm rot} = 70 \pm 10~\rm km\,s^{-1}$, which is consistent with a synchronised system.

Using nine eclipse timings spread over 1100~d, we derived a new linear ephemeris for HS\,2231+2441.  The residuals with respect to this fit do not show any evidence of orbital period variation.

We derived the main parameters of HS\,2231+2441 (see Table \ref{sytem:parameters}) by modelling the B, V, R$_C$, and I$_C$-band light curves and the primary radial velocity curve using the WD code.  Solutions with mass ratio equal to 0.19 and 0.16 were found to be most probable considering the measured photometric and spectroscopic surface gravity of the primary star and the derived mass and radius of the secondary star in comparison with theoretical mass-radius relationship for low mass and brown dwarf stars. Both solutions yield a primary component with low mass (0.19\,M$_{\odot}$ or 0.288\,M$_{\odot}$) and a secondary star being in the brown dwarf regime (0.036\,M$_{\odot}$ or 0.046\,M$_{\odot}$). These masses are in agreement with the results obtained by \citet[][]{Ostensen+2008}. 

The evolutionary status of the primary star in HS\,2231+2441 was analysed by using the effective temperature versus surface gravity diagram. Comparing the primary location in this diagram with post-RGB and post-EHB evolutionary tracks, we concluded that the post-RGB scenario is most likely for the HS\,2231+2441 primary star.

Thus, HS\,2231+2441 has either a ELM white dwarf or a LMWD and brown dwarf star. The secondary star  has one of the lowest masses found among the HW Vir systems \citep[see, e.g.][]{Almeida+2012,Kupfer+2015}. This system may represent the close binaries composed by a low-mass ($<0.5\,M_{\odot}$) WD plus a main-sequence low-mass object discovered in large numbers by \citet[][]{Rebassa-Mansergas+2011}.

\section*{Acknowledgements}
The authors thank the anonymous referee for  helpful suggestions. This study was partially supported by FAPESP (LAA: 2012/09716-6 and 2013/18245-0; AD: 2011/51680-6 CVR: 2013/26258-4), CNPq (CVR: 306701/2015-4) and CAPES. This paper makes use of data obtained from the Isaac Newton Group Archive which is maintained as part of the CASU Astronomical Data Centre at the Institute of Astronomy, Cambridge. The TheoSSA service (http://dc.g-vo.org/theossa) used to retrieve theoretical spectra for this paper was constructed as part of the activities of the German Astrophysical Virtual Observatory.

\end{document}